\documentclass[amsmath,resetfootnote]{aastex702}            

\usepackage{comment}
\usepackage{graphicx}
\usepackage{csquotes}
\usepackage{caption}
\usepackage{booktabs}

\begin{document}

\title{PRICM3: A novel Polarization Ratio framework for three-dimensional CME reconstruction}

\author[0000-0003-2426-2112]{Yara De Leo}
\affiliation{Institute of Physics, University of Graz, A-8010 Graz, Austria}
\email[show]{yara.de-leo@uni-graz.at} 

\author[0000-0003-4867-7558]{Manuela Temmer}
\affiliation{Institute of Physics, University of Graz, A-8010 Graz, Austria}
\email{manuela.temmer@uni-graz.at} 

\author[0000-0002-2047-6327]{Fernando M. L\'opez}
\affiliation{Universidad de Mendoza, CONICET, Grupo de Estudios en Heliofisica de Mendoza, 5500 Mendoza, Argentina}
\affiliation{Observatorio Astron\'omico F\'elix Aguilar (OAFA), Universidad Nacional de San Juan (UNSJ), San Juan, Argentina}
\email{fernando.lopez@um.edu.ar} 

\author[0000-0001-9831-2640]{Sarah E. Gibson}
\affiliation{National Center for Atmospheric Research, Boulder, CO, 80301, USA}
\email{sgibson@ucar.edu} 

\author[0000-0001-7080-2664]{Hebe Cremades}
\affiliation{Universidad de Mendoza, CONICET, Grupo de Estudios en Heliofisica de Mendoza, 5500 Mendoza, Argentina}
\email{hebe.cremades@um.edu.ar} 

\author[0000-0003-0644-1758]{Leonardo Di Lorenzo}
\affiliation{INFAP ``Giorgio Zgrablich'', FCFMyN-UNSL-CONICET, 5700 San Luis, Argentina}
\email{leonardodilorenzo@gmail.com}

\author[0000-0002-1017-7163]{Roberto Susino}
\affiliation{INAF - Torino Astrophysical Observatory, I-10025 Pino Torinese (TO), Italy}
\email{roberto.susino@inaf.it}

\author[0000-0001-9921-1198]{Marco Romoli}
\affiliation{University of Florence - Physics and Astronomy Department, I-50019, Sesto Fiorentino (FI), Italy}
\affiliation{INAF - Arcetri Astrophysical Observatory, I-50125, Florence, Italy}
\email{marco.romoli@unifi.it}

\author[0000-0001-8235-2242]{Lucia Abbo}
\affiliation{INAF - Torino Astrophysical Observatory, I-10025 Pino Torinese (TO), Italy}
\email{lucia.abbo@inaf.it}

\author[0000-0002-8734-808X]{Aleksandr Burtovoi}
\affiliation{University of Florence - Physics and Astronomy Department, I-50019, Sesto Fiorentino (FI), Italy}
\affiliation{INAF - Torino Astrophysical Observatory, I-10025 Pino Torinese (TO), Italy}
\email{aleksandr.burtovoi@inaf.it}

\author[0000-0001-8783-0047]{Chiara Casini}
\affiliation{INAF - Torino Astrophysical Observatory, I-10025 Pino Torinese (TO), Italy}
\email{chiara.casini@inaf.it}

\author[0000-0002-2464-1369]{Michele Fabi}
\affiliation{DiSPeA, University of Urbino Carlo Bo, Urbino (PU), Italy}
\affiliation{INFN, Florence, Italy}
\email{michele.fabi@uniurb.it}

\author[0000-0001-9014-614X]{Federica Frassati}
\affiliation{INAF - Torino Astrophysical Observatory, I-10025 Pino Torinese (TO), Italy}
\email{federica.frassati@inaf.it}

\author[0000-0002-0764-7929]{Giovanna Jerse}
\affiliation{INAF - Astronomical Observatory of Trieste, I-34143, Trieste, Italy}
\email{giovanna.jerse@inaf.it}

\author[0000-0001-8244-9749]{Federico Landini}
\affiliation{INAF - Torino Astrophysical Observatory, I-10025 Pino Torinese (TO), Italy}
\email{federico.landini@inaf.it}

\author[0000-0002-3789-2482]{Maurizio Pancrazzi}
\affiliation{INAF - Torino Astrophysical Observatory, I-10025 Pino Torinese (TO), Italy}
\email{maurizio.pancrazzi@inaf.it}

\author[0000-0002-2433-8706]{Giuliana Russano}
\affiliation{INAF - Astronomical Observatory of Capodimonte, I-80131 Napoli, Italy}
\email{giuliana.russano@inaf.it}

\author[0000-0002-5163-5837]{Clementina Sasso}
\affiliation{INAF - Astronomical Observatory of Capodimonte, I-80131 Napoli, Italy}
\email{clementina.sasso@inaf.it}

\begin{abstract}

Coronal mass ejections (CMEs) are inherently three-dimensional (3D), but traditional white-light coronagraphs provide only two-dimensional (2D) projections. While the Polarization Ratio Technique (PRT) exploits Thomson scattering to infer the line-of-sight (LOS) location of the scattering plasma relative to the plane of the sky, its results are typically represented as a 2D topographical map, in which each image-plane location is associated with an inferred LOS distance. Although these maps encode information on the third spatial dimension, they do not directly provide an explicit 3D representation of the CME geometry, making spatial interpretation and comparison between PRT results from different viewpoints less straightforward.

To address this, we introduce PRICM3 (Polarization Ratio Integrated CME Mapper in 3D), a framework that converts PRT-derived maps into explicit 3D point-cloud reconstructions. This tool enables intuitive visualization, multi-viewpoint comparison, and direct consistency checks against geometric models such as the Graduated Cylindrical Shell (GCS).

As a proof-of-concept, we analyzed a limb CME observed on 28 October 2021 using data from Solar Orbiter/Metis, SOHO/LASCO, and STEREO-A/COR2. The independent reconstructions converged into a consistent 3D volume that aligned remarkably well with the corresponding GCS geometry.

These results prove that the 3D information embedded in PRT topographical maps can be successfully recovered through spatial reconstruction. 
PRICM3 offers a powerful, intuitive method for interpreting polarimetric coronagraph observations, providing a valuable tool for legacy missions such as SOHO and STEREO, for recent ones such as Solar Orbiter, PROBA-3 and PUNCH, as well as for future solar missions.
\end{abstract}


\section{Introduction} \label{sec:Intro}
Coronal Mass Ejections (CMEs) are large-scale eruptions of magnetized plasma from the solar corona that propagate throughout the heliosphere and often interact with planetary environments and spacecraft \citep{Chen:2011,Webb:2012}. They are among the primary drivers of space weather disturbances, including geomagnetic storms, solar energetic particle events, and magnetospheric perturbations \citep{Temmer:2021,Zhang:2021,Gopalswamy:2022}. Despite decades of observations and modeling efforts, important questions remain regarding the three-dimensional (3D) structure of CMEs and their evolution as they propagate away from the Sun. Since white-light (WL) coronagraphs observe Thomson-scattered radiation integrated along the line of sight (LOS), the interpretation of CME morphology is inherently affected by projection effects, making the reconstruction of their true 3D geometry a long-standing challenge.

Several approaches have been developed to recover the 3D structure of CMEs from remote-sensing observations. Stereoscopic techniques such as tie-point triangulation exploit simultaneous observations from different viewpoints to reconstruct the position of specific features in 3D space \citep{Inhester:2006}. Forward-modeling approaches, most notably the Graduated Cylindrical Shell (GCS) model \citep{Thernisien:2006,Thernisien:2009,Thernisien:2011a}, aim at reproducing the global morphology of CMEs by fitting a parameterized flux-rope geometry to multi-viewpoint coronagraph observations. A complementary approach is provided by the Polarization Ratio Technique \citep[PRT; see][]{MoranDavila:2004,Dere:2005,Moran:2010,Mierla:2009,Mierla:2011,deKoning:2011,Temmer:2012, Zuccarello:2013, Mierla:2022}, which exploits the polarization properties of Thomson scattering to infer a proxy for the location of the scattering plasma relative to the plane of the sky (POS), even from observations acquired by a single spacecraft. Polarimetric observations from coronagraphs are routinely provided by SOHO/LASCO \citep{Brueckner:1995} and STEREO/SECCHI \citep{Howard:2008}, and more recently by the Metis coronagraph \citep{Antonucci:2020, Fineschi:2020} aboard Solar Orbiter \citep[SolO;][]{Mueller:2020}. Over the past two decades, the PRT method has therefore become an established approach for investigating the 3D structure of CMEs. 

Traditionally, the PRT encodes 3D CME data into 2D \enquote{topographical maps} via color-coded distance from the POS (e.g., see panels \enquote{D} of the figures in \citet{MoranDavila:2004}, or Fig. 18-20 in \cite{Zuccarello:2013}).   
While valuable for analyzing morphology and kinematics, this representation is inherently indirect: the inferred third spatial coordinate is encoded as a value associated with each position in the 2D image plane, rather than being represented explicitly in three-dimensional space. Consequently, the three-dimensional geometry of the CME must be interpreted from the topographical map rather than directly visualized.

%
Because different coronagraphs differ in their fields of view, pixel scales, observing geometries, and acquisition cadences for total ($tB$) and polarized brightness ($pB$) measurements, current polarimetric datasets suffer from inconsistent spatial and temporal sampling. This lack of agreement severely impedes the alignment and joint analysis of multi-spacecraft observations within a unified 3D framework, underscoring a critical gap in the field: the absence of a general-purpose tool for mapping multi-viewpoint polarimetric data as/into 3D spatial reconstructions. We address this gap by providing a framework, PRICM3 (Polarization Ratio Integrated CME Mapper in 3D), transforming PRT-derived depth information into intuitive and physically meaningful 3D representations. PRICM3 is a multi-spacecraft approach enabling improved visualization, cross-viewpoint comparison, and validation against independent reconstruction methods.
Through these capabilities, the framework enables scientists to map the 3D envelope of a CME, investigate its 3D evolution, and derive robust coronal diagnostics.
The continued expansion of space-based polarimetric coronagraph observations makes the development of tools such as PRICM3 timely, enabling the scientific community to fully exploit and jointly analyze multi-spacecraft polarimetric observations. In particular, polarimetric data products from PROBA-3/ASPIICS \citep{Zhukov:2026} and PUNCH \citep[][]{2026SoPh..301...16D}, which are expected to become available on a continuous basis, will further expand the scientific potential of PRT-based studies. 

In this work, we present the capabilities of PRICM3 and its application on an example of stereoscopic PRT diagnostics using SOHO, SolO and STEREO-A data. We elaborate on how to convert PRT topographical maps into 3D point-cloud reconstructions of a CME within a common heliocentric coordinate system (Heliocentric Earth Equatorial, HEEQ\footnote{HEEQ is a Sun-centered Cartesian coordinate system whose $Z$-axis is aligned with the solar rotation axis, ``while the $X$-axis is the intersection between solar equator and solar central meridian of date'' \citep[taken from][]{Franz:2002}.}).
Throughout this work, we use Cartesian HEEQ coordinates for the 3D representation of the reconstructed structures and the corresponding Stonyhurst heliographic coordinates to express their angular positions. The method enables a direct comparison between independent PRT reconstructions obtained from different observing geometries and facilitates their comparison with GCS-derived CME appearance.
\begin{figure*}
\centering
\includegraphics[width=0.98\textwidth]{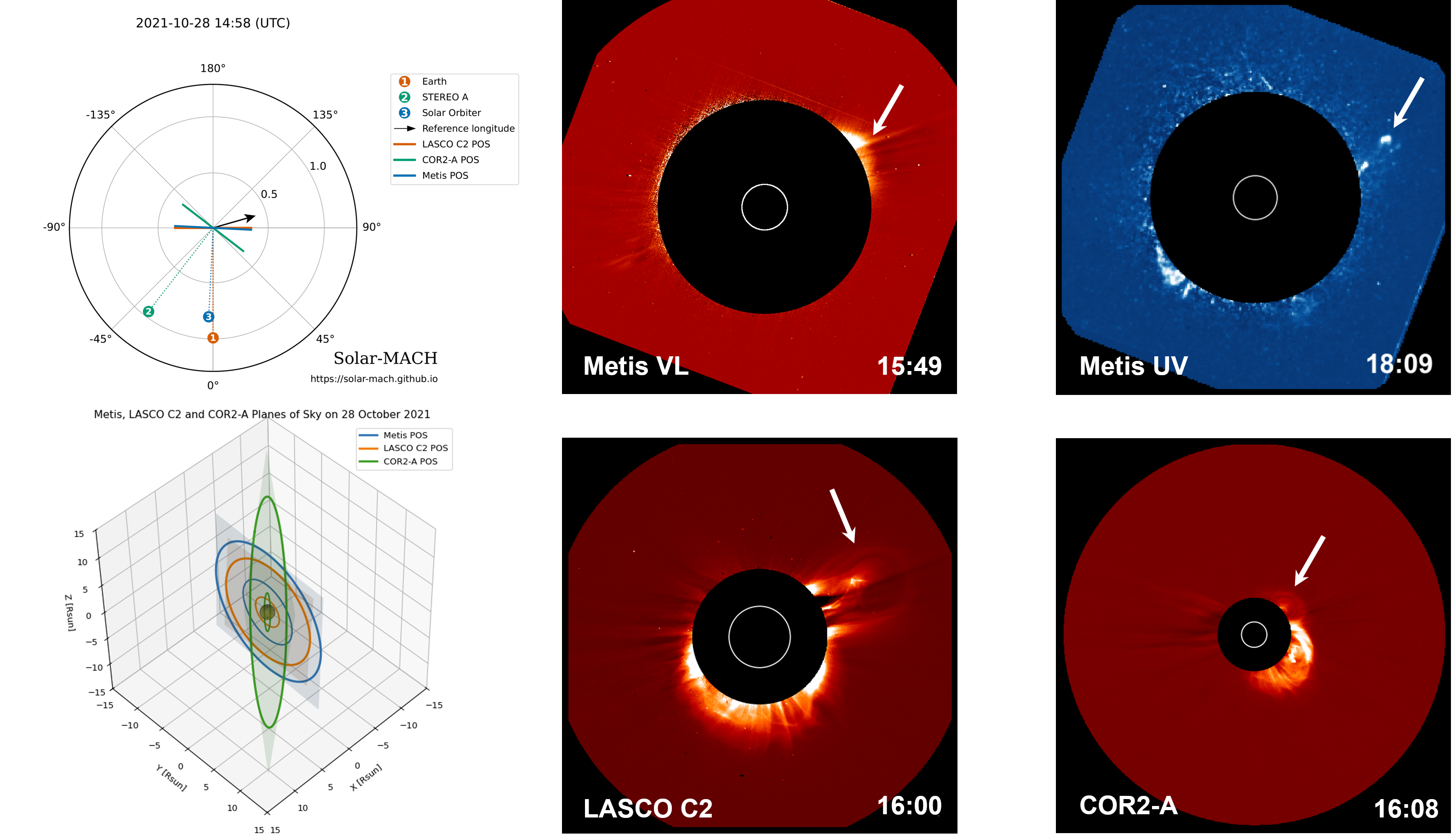}
\caption{Observational overview of the 28 October 2021 limb CME and viewing geometry. \textit{Top:} The left panel represents the Spacecraft constellation obtained in Stonyhurst coordinates, adapted from Solar-MACH \citep{Gieseler:2022}, showing the relative positions of Solar Orbiter, Earth, and STEREO-A and their planes of sky. The black arrow indicates the propagation direction of the limb CME. The other two panels show the event in a Metis $tB$ image (at 15:49 UT) and UV one (at 18:09 UT), respectively. \textit{Bottom:} The left panel shows a screenshot of the interactive 3D visualization of the observing geometry generated with PRICM3. The colored planes indicate the POS of Metis (blue), LASCO C2 (orange), and STEREO-A/COR2 (green), while the corresponding circles mark the inner and outer limits of their fields of view. The other two panels show the event in $tB$ images acquired by LASCO C2 (at 16:00 UT) and COR2-A (at 16:08 UT), respectively.} 
    \label{fig:event}%
\end{figure*}
Section \ref{sec:data_meth} describes the event study of the CME observed on 28 October 2021 (see De Leo et al., 2026b, \textit{in preparation}, for further details) and the methodology adopted in this work.
Section \ref{sec:PRICM3} introduces the PRICM3 framework, describing its workflow for converting PRT topographical maps into 3D point clouds and its integration with the GCS model for quantitative consistency assessment. Section \ref{sec:results} presents the application of PRICM3 to the selected multi-viewpoint CME event, including the resulting 3D reconstructions and their quantitative comparison with the GCS model. Finally, Section \ref{sec:conclusions} discusses the main implications and limitations of our results, explores potential applications to new polarimetric datasets, and summarizes the conclusions of this work.
\section{Data and Methodology}\label{sec:data_meth}
During its six years of operation, the SolO/Metis coronagraph has demonstrated its capability to routinely detect and characterize CMEs across heliocentric distances ranging approximately from 1.7 to 9 $R_\odot$, depending on the spacecraft-Sun distance, providing new insights into their morphology and physical properties \citep{Andretta:2021,Bemporad:2022,Bemporad:2024,Bemporad:2025,Zimbardo:2023,Frassati:2024,Russano:2024,Russano:2026}. Beyond its unique capability to simultaneously image the extended corona in visible light (VL) and in the ultraviolet (UV) H I Lyman-$\alpha$ emission (including observations from out-of-ecliptic perspectives), Metis is particularly well suited for the application of the PRT because nearly all of its observing programs provide cotemporal $pB$ and $tB$ measurements at variable cadences tailored to different scientific objectives. The only exception is a high-cadence observing program designed to acquire $tB$ images every 20 s, for which the acquisition of the four exposures required to derive $pB$ is not compatible with the targeted cadence. Polarimetric observations are also routinely provided by other space-based coronagraphs, although with different observing strategies and cadences. SOHO/LASCO C2 typically acquires polarized sequences three times per day, approximately every 8 hr, while LASCO C3 generally provides one to two polarized sequences per day. The STEREO/SECCHI coronagraphs provide polarized observations at higher cadences, with COR1-A and COR2-A typically acquiring polarimetric sequences approximately every 5 min and 1 hr, respectively.
\subsection{Event and observations}\label{sec:event}
As a proof of concept, we apply PRICM3 to a relatively slow CME (GCS-derived apex speed of $\sim$350 km/s) observed near the northwestern solar limb on 28 October 2021\footnote{Later the same day, a fast halo CME associated with the first X-class flare of Solar Cycle 25 erupted from NOAA AR 12887. The present study focuses exclusively on the earlier slow limb CME.}, during Solar Orbiter’s cruise phase (see Fig.~\ref{fig:event}). Previously analyzed by \citet{Russano:2024} as one of six CMEs, this event was selected for its favorable geometry, with propagation close to the POS from Earth and Metis, minimizing projection effects and facilitating morphological analyses. The event exhibited the classical three-part morphology commonly observed in WL coronagraph images \citep{Illing:1985}.

At the time of the eruption, Solar Orbiter was located at approximately 0.8 au and nearly aligned with the Sun--Earth line, with a longitudinal separation of less than 3$^\circ$. Consequently, Metis and the near-Earth coronagraph LASCO C2 observed the CME from nearly the same perspective. In contrast, STEREO-A was positioned approximately $-$38$^\circ$ away in Stonyhurst longitude, providing an independent viewing geometry particularly suitable for 3D reconstruction and validation. The spacecraft constellation for this date is shown in the top-left panel of Fig. \ref{fig:event}.

During the 28 October 2021 synoptic campaign, the Metis field of view (FOV) extended from about 4.8 to 10.3 $R_{\odot}$. We used Level 2 UV, $pB$, and $tB$ observations from the Data Release 1.0 (DR1), radiometrically calibrated as described in \citet{De_Leo:2023,De_Leo:2025}. During the observing sequence, Metis acquired $pB$ and $tB$ images at a cadence of approximately 30 minutes, while the UV observations had a cadence of 2 minutes. 
The Metis observation timestamps refer to the Solar Orbiter observing frame. To compare these measurements with observations acquired from near Earth, we account for the approximately 1.67 min difference in light-travel time between Solar Orbiter and Earth.
%
The VL and UV channels have nominal pixel scales of
10.7 and 20 arcsec/pixel, respectively, at full resolution.
For the observing sequence analyzed here, the images
were acquired with $2\times2$ and $4\times4$ on-board
binning, resulting in effective angular pixel scales of
21.4 and 80 arcsec/pixel, respectively.
The corresponding projected physical
pixel sizes, which depend on the heliocentric distance
of Solar Orbiter, were approximately $11.7\times10^3$
and $47.1\times10^3$ km/pixel at the time of the
observations.

The limb CME first appeared in the STEREO-A/COR1 FOV at 13:56 UT, entered the COR2 FOV at 15:08 UT, and remained detectable until approximately 21:00 UT. 
%
From the near-Earth perspective, the event was observed by LASCO C2 from 14:00 to 18:24 UT and subsequently crossed the Metis FOV between approximately 15:30 UT and 19:40 UT.
The raw COR2 data (Level 0) were downloaded from the Virtual Solar Observatory (VSO)\footnote{\url{https://sdac.virtualsolar.org/cgi/search}}, and processed using the standard \texttt{secchi\_prep} routine available in SolarSoft. The Level 1 polarized LASCO C2 observations were obtained from the dedicated LASCO polarization database \footnote{\url{https://lasco-www.nrl.navy.mil/lz/polarize/}}. The main observational characteristics of the three coronagraphs, including their FOVs, polarimetric cadences, and pixel scales, are summarized in Table~\ref{tab:instrument_characteristics}. 
\begin{table}[ht]
\centering
\caption{Main observational characteristics of the coronagraphs
used in this study. Pixel scales refer to the native detector
sampling, before on-board binning.}
\label{tab:instrument_characteristics}
\begin{tabular}{lccc}
\hline\hline
Instrument &
FOV ($R_{\odot}$) &
pB cadence &
Pixel scale ($''$ pixel$^{-1}$) \\
\hline
Metis VL   & $\sim$4.8--10.3$^{a}$ & $\sim$30 min & 10.7 \\
LASCO C2   & 2.2--8.0           & $\sim$8 h    & 11.9 \\
COR2-A     & 2.5--15.0          & $\sim$1 h    & 14.7 \\
\hline
\end{tabular}
\smallskip
\begin{minipage}{0.95\columnwidth}
\footnotesize
$^{a}$ The Metis FOV depends on the heliocentric
distance of Solar Orbiter. The reported range
represents the approximate coverage over its orbit.
The Metis cadence refers to the observing sequence
analyzed here; the LASCO C2 and COR2-A values
are typical polarimetric cadences.
\end{minipage}
\end{table}
\begin{figure*}
\centering
\includegraphics[width=0.98\textwidth]{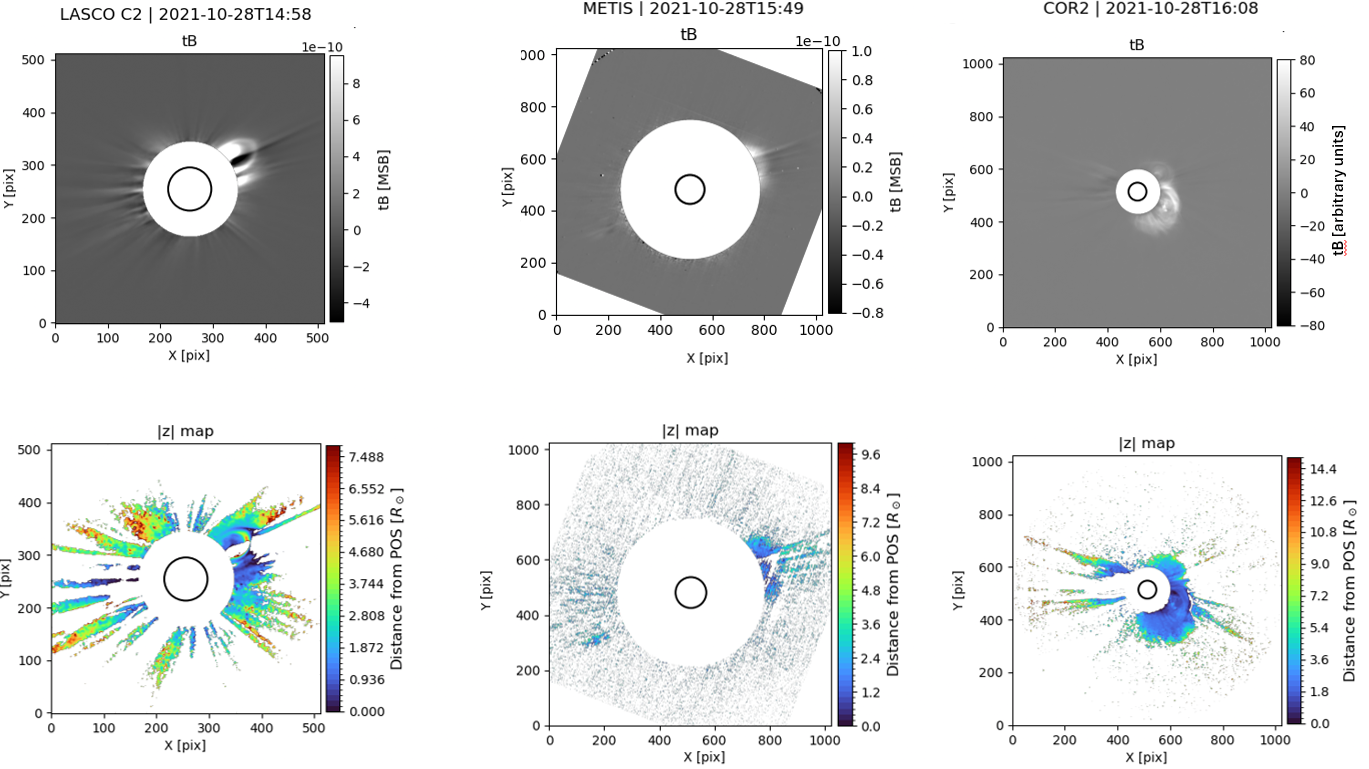}
\caption{Pre-event subtracted total brightness images ($tB$ in top panels) and corresponding topographical maps derived from the PRT ($|z|$ maps in bottom panels) for different observing geometries and times. The color scale encodes the absolute distance, $|z|$, from the observer's POS, expressed in solar radii. From left to right, the three columns show the CME observed by LASCO C2 (14:58 UT), Metis (15:49 UT), and COR2-A (16:08 UT) on 28 October 2021, respectively.} 
    \label{fig:top_maps}%
\end{figure*}
\subsection{Derivation of the PRT topographical maps}
PRICM3 is based on PRT depth information, which is then used to infer the 3D spatial representation of a CME and to compare results across multiple spacecraft viewpoints. Figure~\ref{fig:top_maps} shows the PRT topographical maps derived from polarized WL coronagraph observations of the CME from three different viewpoints, providing estimates of the scattering plasma distances, $|z|$, along the LOS relative to the corresponding POS. For context the top row of Fig.~\ref{fig:top_maps} presents the corresponding background-subtracted $tB$ images. We note that the different polarimetric cadences of the three coronagraphs resulted in observations being acquired at different times, limiting the availability of near-simultaneous datasets and complicating the direct comparison of CME structures reconstructed from the three viewpoints. The procedure used to derive the PRT topographical maps, together with the pre-event coronal configuration providing context for background structures in the region of interest, is described and illustrated in Appendix~\ref{appendix:a}.
\section{PRICM3: Polarization Ratio Integrated CME Mapper in 3D}\label{sec:PRICM3}
\begin{figure*}
\centering
\includegraphics[width=0.74\textwidth]{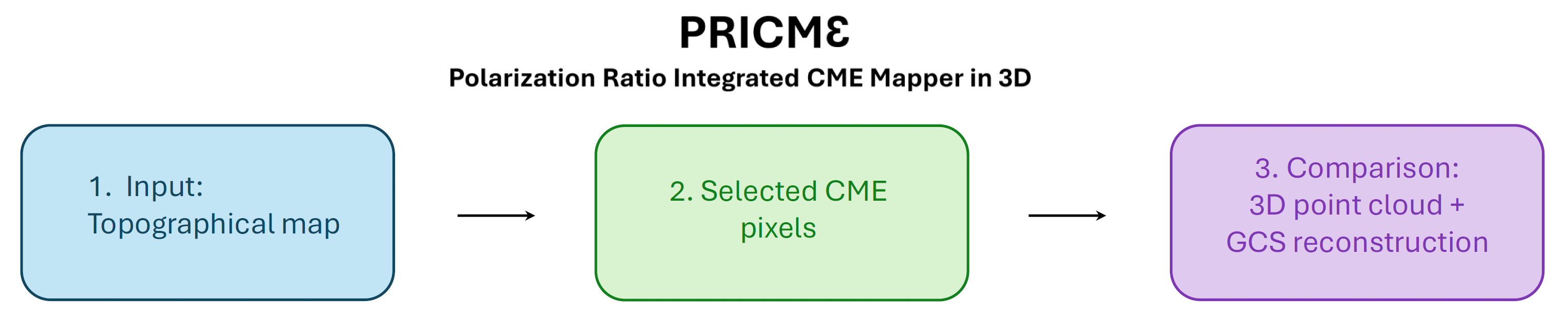}\\
\includegraphics[width=0.99\textwidth]{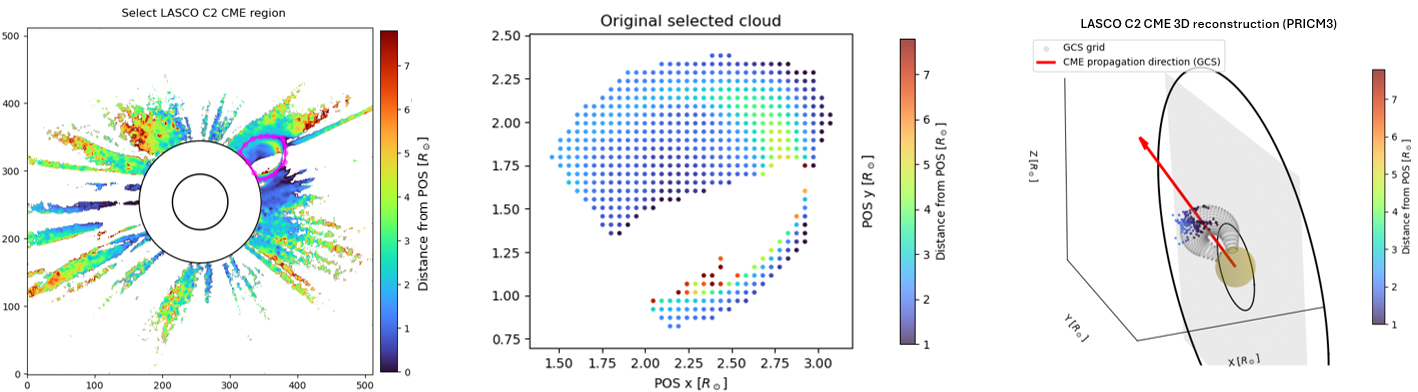}\\
\includegraphics[width=1.0\textwidth]{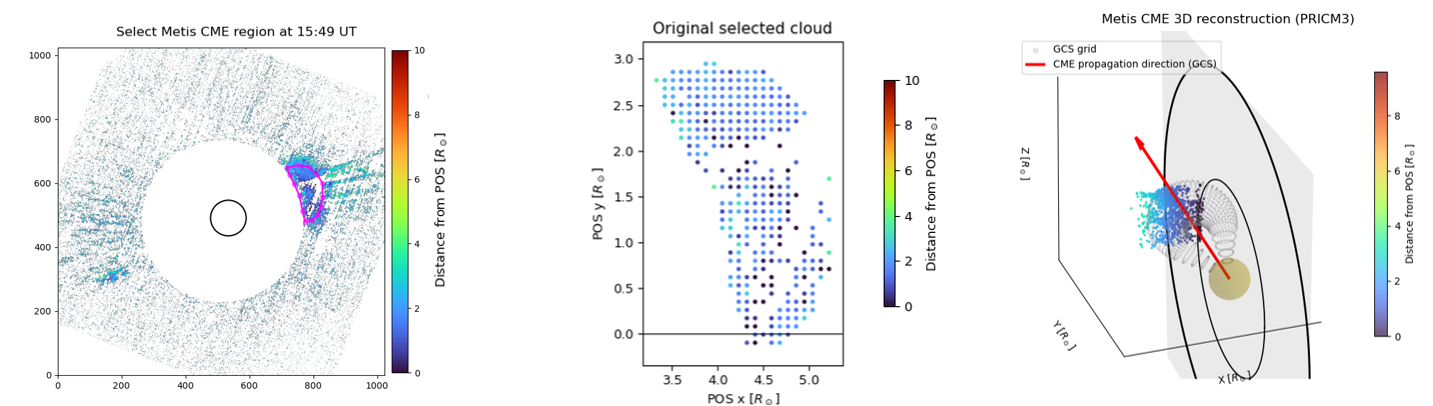}\\
\includegraphics[width=0.99\textwidth]{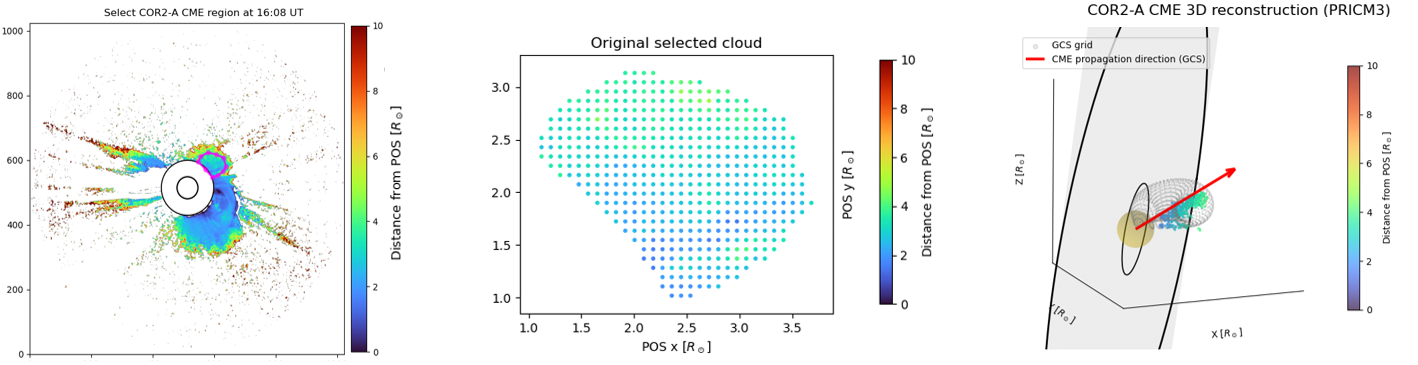}\\
\caption{Top: Schematic representation of the PRICM3 workflow. Bottom: Illustration of the main steps of the methodology applied to LASCO C2 (first row), Metis (second row), and COR2-A (third row) observations of the 28 October 2021 CME, at the same timestamps as in Fig. \ref{fig:top_maps}. From left to right, the column panels show the selection of the ROI (outlined in magenta) from the input topographical map, the corresponding point cloud displayed in the observer's POS coordinate system, and its 3D reconstruction in a common Sun-centered reference frame, with the GCS wireframe superimposed and the CME propagation direction inferred from it (red arrow). The point-cloud sampling density is user-configurable and has been reduced here for visualization purposes to improve the clarity of the plots.}
\label{fig:workflow}
\end{figure*}

The PRICM3 workflow is presented in the top panel of Figure~\ref{fig:workflow}. PRICM3 takes as its primary input the precomputed PRT topographical maps in FITS format (see details in Appendix~\ref{appendix:a}). To transform these maps into 3D point clouds, PRICM3 requires the observing geometry, which can be retrieved directly from the topographical map headers, provided that the relevant metadata from the original observations have been preserved. Otherwise, the corresponding $tB$ FITS image headers are used to retrieve the necessary geometric information. 
%
Based on the observational metadata, the position of
each observer and the orientation of the corresponding
POS are determined, so that the PRICM3 reconstructed
structures are presented in a common heliocentric
coordinate system (HEEQ).
Upon transformation from the local observer frame
to the HEEQ reference frame, the LOS coordinate $z$
contributes to the Cartesian coordinates $X$, $Y$, and $Z$
of the reconstructed point cloud, depending on the
heliographic longitude and latitude of the observer
(see the right panels of Fig.~\ref{fig:workflow}).

A crucial step of the procedure is the conversion of the inferred depth information from the topographical maps into 3D point clouds specifically isolating the region of interest (ROI), i.e., the CME. For this, PRICM3 provides an interactive selector to choose that ROI as accurately as possible. The projected coordinates of each interactively selected CME pixel are combined with the reconstructed LOS distance inferred from the PRT analysis to determine the 3D location of the scattering plasma in the local observer frame. Finally, these local coordinates are subsequently transformed from the local POS reference frame into the common heliocentric coordinate system, allowing direct spatial comparison between reconstructions obtained from different viewpoints. In this regard, the corresponding $tB$ images provide important complementary information for identifying CME morphology and assessing potential contamination from background coronal structures.

Since PRT topographical maps do not preserve the observed brightness distribution, distinguishing CME features from unrelated coronal structures based on inferred LOS distances alone can be challenging. To refine the reconstructed CME structures, PRICM3 provides two filtering options: a statistical filter based on the 95\% confidence interval (CI) of the $|z|$ distribution to exclude extreme LOS-distance values, and a user-defined $|z|$ cutoff to remove structures deemed unrelated to the CME based on their morphology and spatial distribution. A detailed step-by-step procedure applying PRICM3 is presented in Section~\ref{sec:results}.
\subsection{GCS Model Integration and Quantitative Consistency Assessment}\label{sec:GCS}
The GCS model provides a large-scale representation of the CME flux-rope geometry, whereas PRT point clouds represent plasma locations inferred from polarimetric observations. To investigate the geometrical consistency between these complementary representations, PRICM3 allows users to incorporate GCS geometries obtained independently through the GCS forward-modeling technique. The GCS fitting itself is not performed within the PRICM3 framework; instead, users may provide parameters derived from an external fitting procedure \citep{Thernisien:2006,Thernisien:2009}, which PRICM3 uses to generate the 3D mesh representing the characteristic \enquote{hollow croissant} flux-rope geometry. The resulting mesh is displayed alongside the PRT point clouds in a common heliocentric reference frame, enabling direct visual comparison of the reconstructed plasma distribution with the large-scale CME morphology predicted by the GCS model.

To assess the geometrical consistency between the reconstructed PRT point clouds and the GCS model, PRICM3 provides two complementary quantitative estimators. The first estimator evaluates the angular offsets between the centroid of each reconstructed point cloud and the CME propagation direction inferred from the GCS model, expressed in terms of Stonyhurst longitude and latitude. The second estimator computes the fraction of reconstructed points enclosed within the GCS volume, providing a measure of the spatial correspondence between the PRT-derived plasma distribution and the independently modeled CME geometry. These estimators characterize different aspects of the reconstruction: the first assesses the alignment of the point-cloud centroid with the inferred CME propagation direction, whereas the second evaluates the spatial distribution of the reconstructed points relative to the GCS envelope. Consequently, a small angular offset does not necessarily imply a high fraction of points enclosed within the GCS volume.

To quantitatively compare and evaluate the geometrical agreement between the reconstructed point clouds and the GCS model, uncertainties associated with both the PRT reconstruction and the GCS model are considered. For the reconstructed point clouds, a typical uncertainty of 3\% was assumed for the local LOS coordinate $z$ \citep{Lopez:2012, Zuccarello:2013}. For the GCS reconstruction, uncertainties of $22^\circ$ in Stonyhurst longitude, $4^\circ$ in latitude, and 5\% in apex height were adopted following \citet{Verbeke:2023}.
The final output of PRICM3 is a 3D representation of the CME that simultaneously displays independent PRT reconstructions from different instruments and viewpoints, the spacecraft observing geometry, and, when provided, the corresponding GCS shell. This integrated visualization facilitates the assessment of the spatial consistency of multi-viewpoint reconstructions beyond the limitations of conventional 2D topographical maps. Furthermore, it offers novel diagnostic capabilities for both the 3D CME body and its surrounding envelope. PRICM3 is openly available to the community through a Zenodo repository\footnote{PRICM3 (De Leo et al., 2026), Zenodo, DOI: 10.5281/zenodo.23734905. The repository is currently under restricted access and will be made publicly available upon acceptance of this manuscript.} and supports the export of reconstructed 3D point clouds in JSON format for further analysis and visualization using external tools, including modified versions of JHelioviewer\footnote{HelioFITS Studio (Version v0.9.2), developed by C. R. Gilly (2026), available on Zenodo at \url{https://doi.org/10.5281/zenodo.24177191}}.
\section{Application of PRICM3 to a Multi-Viewpoint CME Event} \label{sec:results}
To demonstrate the practical application of PRICM3, we apply the framework to the slow limb CME observed on 28 October 2021, as observed by Metis, LASCO C2, and COR2-A (see Section~\ref{sec:event}). As shown in the last three rows of panels in Fig.~\ref{fig:workflow}, we exercise through the workflow starting from PRT topographical maps to the final 3D point-cloud reconstruction, as well as demonstrate the capabilities of PRICM3 for visualizing and comparing CME structures within a common heliocentric reference frame. 

\begin{figure}[t]
\centering
\includegraphics[width=0.9\columnwidth]{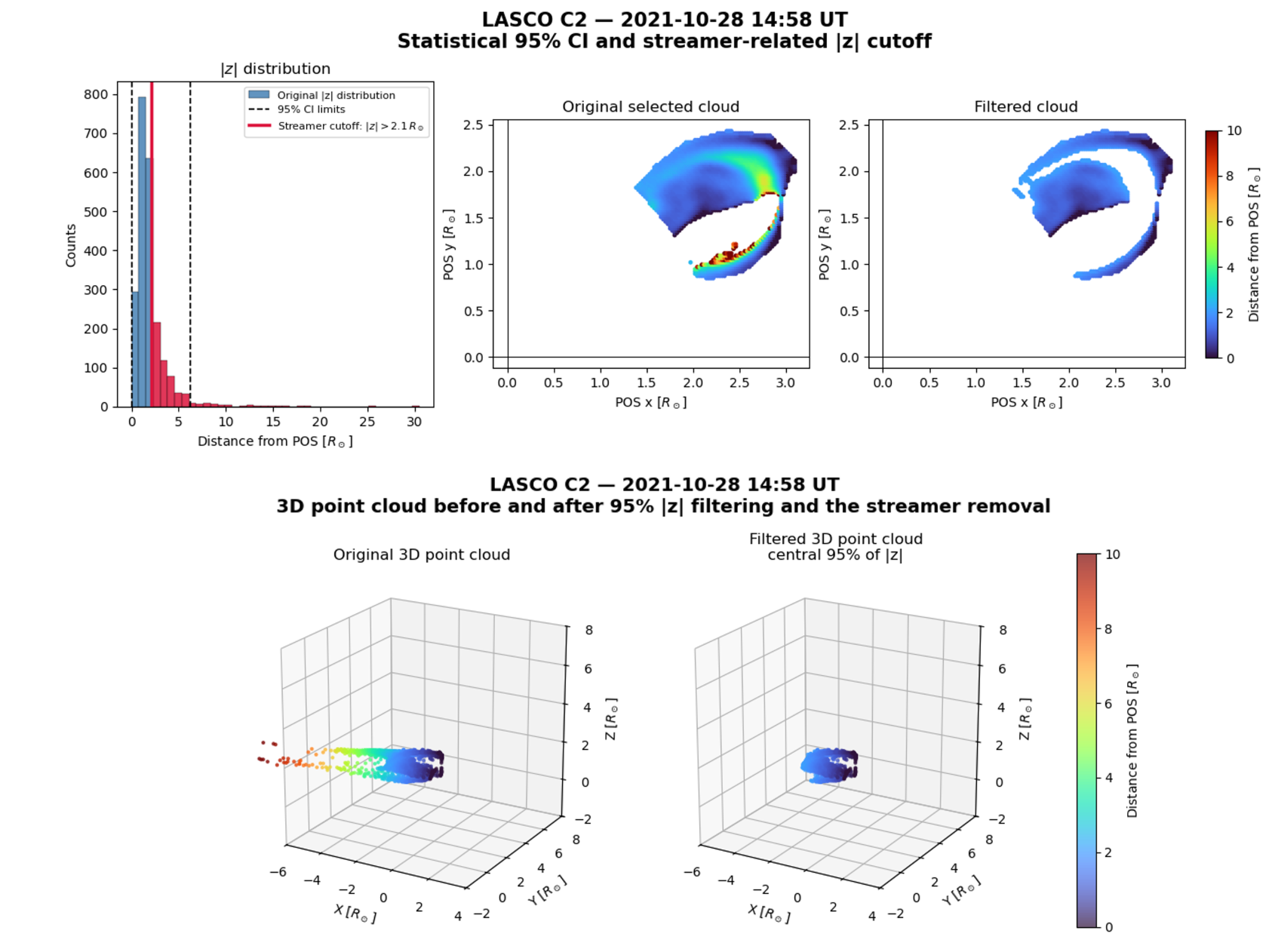}
\caption{
Filtering procedure adopted to isolate the CME plasma from the LASCO C2 reconstruction at 14:58~UT. The upper panels illustrate the statistical and morphology-based filtering applied to the reconstructed $|z|$ distribution and topographical map, while the lower panels show the corresponding 3D point clouds before and after filtering.
}
\label{fig:appendix_filtering}
\end{figure}

Following the preliminary brightness-based filtering applied during the derivation of the PRT topographical maps (Appendix~\ref{appendix:a}), the CME region was further isolated, to exclude structures interpreted as not belonging to the CME, e.g., streamers. The accuracy of the resulting 3D point clouds critically depends on the quality of the reconstructed CME signal. Hence, particular care needs to be devoted to distinguishing the CME from surrounding coronal structures. A particularly illustrative example to present the complete procedure for isolating the CME signal (i.e., deriving the filtered cloud) is shown in Fig.~\ref{fig:appendix_filtering}, provided by the LASCO C2 reconstruction at 14:58 UT  -- the only timestamp available during the event.

The procedure begins with the interactive selection of the CME Region of Interest (ROI) from the input topographical map, as indicated by the magenta contour in the left panels of Fig.~\ref{fig:workflow}. Subsequently  statistical filtering is applied to retain the 95\% CI of the reconstructed $|z|$ distribution (top left panel of Fig.~\ref{fig:appendix_filtering}), from which the original selected cloud is derived (top middle panel of Fig.~\ref{fig:appendix_filtering}). When compared to the $tB$ image (top left panel of Fig.~\ref{fig:top_maps}), which exhibits the characteristic CME three-part morphology, the selected region in the topographical map reveals a distinct crescent-shaped feature highlighted in green shades approximately 4--6\,R$_\odot$ away from the POS. This feature matches the dark void observed in the $tB$ image. Consequently, the pixels shown in green shades correspond to plasma belonging to other structures; we interpret this feature as part of the streamer located at a position angle of approximately 300$^\circ$, viewed through the CME cavity along the LOS.

Recognizing the physical origin of this structure -- an insight uniquely enabled by the topographical maps -- allowed us to fully isolate the CME plasma from unrelated coronal emission. To achieve this, a morphology-based cutoff was performed (specifically, for $|z| \geq 2.1\,R_\odot$, done iteratively to best suppress the streamer-related structure without affecting the main body of the eruption). Through this process, the final filtered cloud was successfully derived (top-right panel of Fig.~\ref{fig:appendix_filtering}).

\begin{figure*}
\centering
\includegraphics[width=0.98\textwidth]{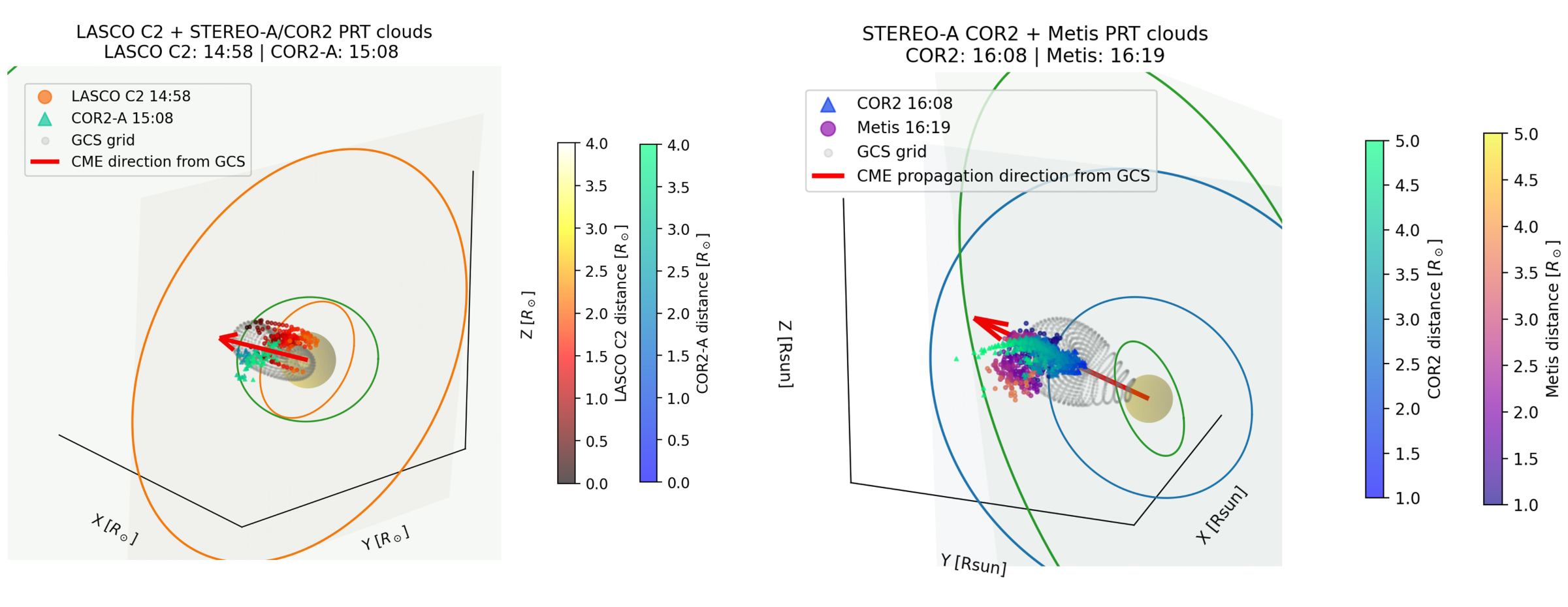}
\caption{Comparison between independent PRT point-cloud reconstructions of the 28 October 2021 CME and the corresponding GCS forward-model reconstruction. The left panel compares the LASCO C2 (14:58 UT) and COR2-A (15:08 UT) reconstructions, while the right panel shows the COR2-A (16:08 UT) and Metis (16:19 UT) reconstructions. In both panels, only the GCS mesh corresponding to the latest observation is displayed for clarity, together with the inferred CME propagation direction. Point colors indicate the distance from the plane of the sky. For visualization purposes, only a randomly selected subset of each reconstructed point cloud is shown, whereas all quantitative analyses presented in this work were performed using the complete point clouds. An accompanying movie, available as supplementary material, shows both 3D views rotating around the vertical axis, allowing the spatial distribution of the reconstructed point clouds and their relative positions with respect to the GCS model to be appreciated from different viewing angles.}
    \label{fig:results}%
\end{figure*}

Finally, these data were transformed into a 3D point cloud by assigning each pixel its projected position in the local POS and the corresponding LOS distance inferred from the PRT analysis. The lower panels of Fig.~\ref{fig:appendix_filtering} compare the 3D point cloud derived from the original selected cloud versus the filtered cloud, illustrating the 3D PRT-related information of the pixels and stressing the overall importance of the filtering procedure. This filtering methodology was consistently adopted for all topographical maps analyzed in this work.
Figure~\ref{fig:results} presents two representative comparisons between independent PRT reconstructions of the event with the closest observations in time. The left panel shows the reconstructions derived from LASCO C2 at 14:58~UT and COR2-A at 15:08~UT, while the right panel compares the COR2-A reconstruction at 16:08~UT with the Metis reconstruction at 16:19~UT. 
Despite the different observing geometries, instrumental characteristics, and completely independent applications of the PRT, both comparisons yield coherent 3D CME structures. On visual inspection, we see a much better agreement between the COR2-A and Metis reconstructions compared to LASCO C2 and COR2-A, reflecting the different observing conditions and reconstruction uncertainties. 

To demonstrate and assess the geometrical consistency derived from an independent geometrical reference, the GCS reconstruction is superimposed as a 3D mesh together with the inferred CME propagation direction (Stonyhurst longitude $106^\circ$, latitude $25^\circ$). The GCS forward modeling was performed outside the PRICM3 framework, using LASCO C2 clear-filter and COR2-A double-exposure observations ($tB$), whose acquisition times differ from the polarized measurements employed for the PRT reconstructions. Assuming a self-similar expansion of the CME over the time interval considered, a linear fit to the GCS apex height as a function of time was used to interpolate the model apex height to the timestamps analyzed in this work with the PRT. The resulting GCS parameters, including the interpolated apex heights, were then provided as input to PRICM3 to visualize the corresponding 3D mesh for comparison with the PRT point clouds (see also Section~\ref{sec:GCS}).

\begin{figure}[t]
\centering
\includegraphics[width=0.8\columnwidth]{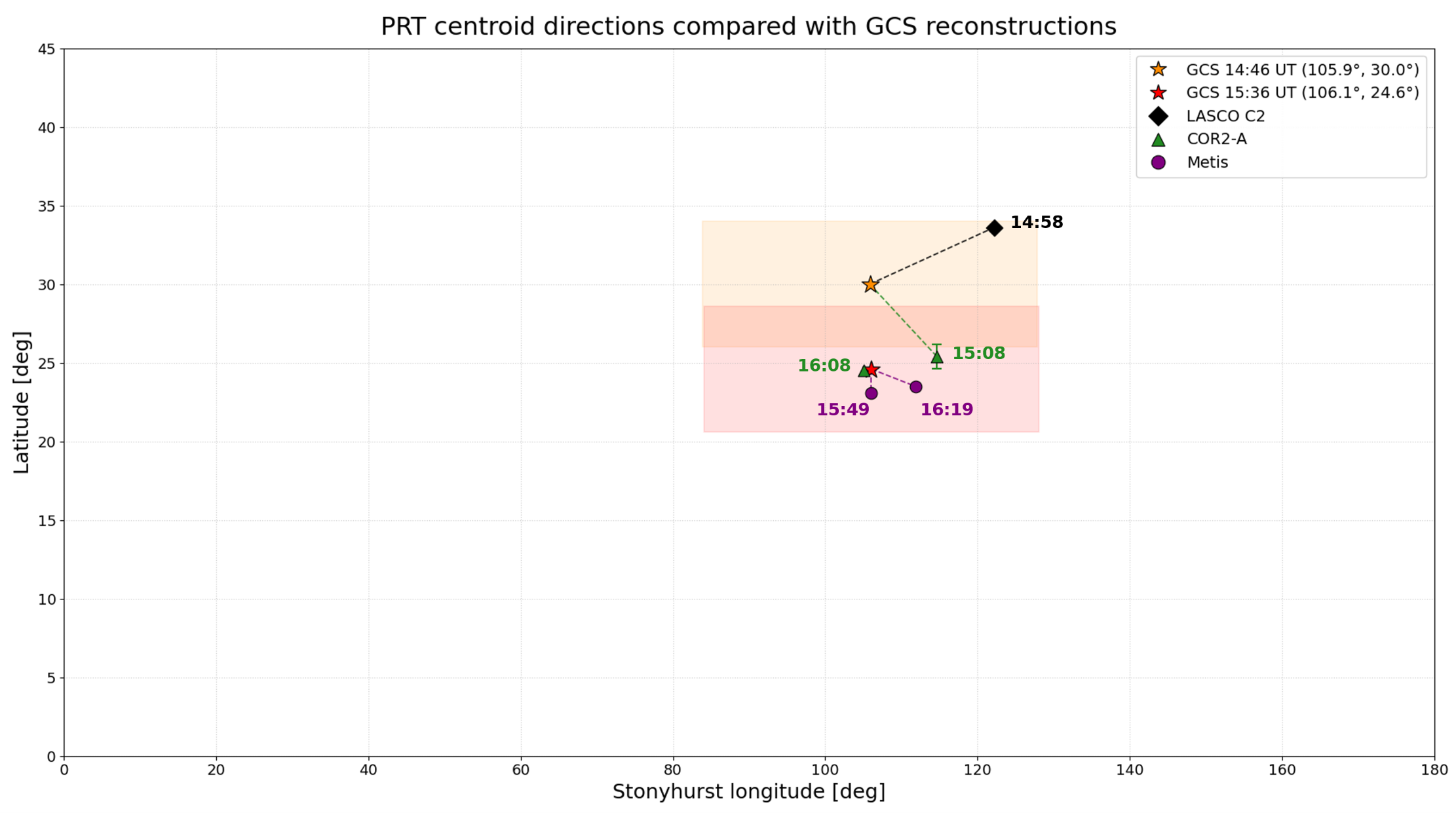}
\caption{Comparison between the reconstructed PRT cloud centroids and the CME propagation directions inferred from the GCS model. The longitude and latitude axes are intentionally shown over a broad range to provide geometrical context, although the reconstructed centroids occupy only a small region of the plot (see the main text for details).}
\label{fig:centroids}
\end{figure}

We see that for COR2-A (16:08 UT) and Metis (16:19 UT) most of the reconstructed points are contained within the GCS volume and preferentially distributed around the inferred propagation axis (see right panel of Fig.~\ref{fig:results}). A similarly good correspondence with the GCS model is found for the Metis reconstruction at 15:49~UT (shown in Appendix \ref{appendix:b}). In contrast, the earlier LASCO C2 and COR2-A reconstructions exhibit a lower degree of mutual agreement. While the LASCO point cloud remains broadly consistent with the GCS envelope, the corresponding COR2-A reconstruction shows larger deviations from the model geometry. This lower level of agreement is readily explained by the early evolutionary stage of the eruption in the COR2-A observation. At 15:08~UT the CME was still barely emerging above the COR2-A occulter, providing only limited observational constraints for both the PRT reconstruction and the GCS modelling. The visual inspection is complemented by deriving the two quantitative estimators implemented in PRICM3 and described in Sect.~\ref{sec:GCS}.
The results obtained using the first estimator to assess the alignment of the point-cloud centroid with the inferred CME propagation direction, are presented in Fig.~\ref{fig:centroids}. It shows the centroid locations in Stonyhurst longitude and latitude, with the black diamond, green triangles, and purple circles representing the centroids of the LASCO C2, COR2-A, and Metis point clouds, respectively, while the orange and red stars indicate the GCS propagation directions corresponding to the early (14:46 UT) and late (15:36 UT) reconstructions. The shaded regions represent the adopted GCS uncertainties of ±22° in longitude and ±4° in latitude, whereas the dashed lines connect each reconstructed point cloud to the corresponding GCS solution. All reconstructed centroids remain within these uncertainty bounds, indicating that the independently derived PRT reconstructions consistently recover the large-scale CME propagation direction.
The second estimator evaluates the geometrical agreement between the reconstructed PRT point clouds and the GCS model by computing the fraction of reconstructed points enclosed within the GCS volume. To account for GCS reconstruction uncertainties, three values of the GCS apex height were considered: the nominal value and variations of $\pm5\%$. Similarly, three configurations of each PRT point cloud were evaluated, corresponding to the nominal reconstructed $|z|$ values and variations of $\pm3\%$. The overlap fraction was calculated for each combination of GCS apex height and PRT LOS-distance values, and the maximum value was retained as the maximum overlap. This quantity therefore represents the most favorable geometrical agreement within the adopted uncertainty ranges, rather than the overlap obtained from the nominal reconstructions alone.
The highest overlap fractions are found for the cases of COR2-A (16:08 UT) and Metis (16:19 UT), with $\sim$86\% and $\sim$74\%, respectively,  and for the Metis reconstruction at 15:49, with $\sim$87\%. As already discussed, the earliest COR2 reconstruction yields the smallest overlap because the CME is only beginning to emerge above the occulter. The results of the quantitative comparison using both estimators are summarized in Table~\ref{tab:validation}.

\begin{table}
\centering
\caption{Summary of the quantitative comparison between the reconstructed PRT point clouds and the GCS model. The quantities $\Delta$Lon and $\Delta$Lat, obtained from the first estimator, denote the angular offsets between the reconstructed point-cloud centroids and the GCS propagation direction, while the last column reports the maximum fraction of reconstructed points enclosed within the GCS volume after accounting for the adopted uncertainties. See the main text for details.}
\label{tab:validation}
\small
\begin{tabular}{lccccc}
\toprule
Instrument & Time (UT) &
Nominal GCS apex ($R_\odot$) &
$\Delta$Lon ($^\circ$) &
$\Delta$Lat ($^\circ$) &
Max. overlap (\%) \\
\midrule
LASCO C2 & 14:58 & $3.70 \pm 0.19$ & $16.3^\circ$ & $3.6^\circ$ & 68.1 \\
COR2-A   & 15:08 & $3.90 \pm 0.20$ & $8.8^\circ$ & $-4.6^\circ$ & 16.2 \\
Metis    & 15:49 & $5.70 \pm 0.29$ & $0.0^\circ$ & $-1.5^\circ$ & 86.9 \\
COR2-A   & 16:08 & $5.58 \pm 0.28$ & $-1.0^\circ$ & $-0.1^\circ$ & 85.9 \\
Metis    & 16:19 & $6.20 \pm 0.31$ & $5.8^\circ$ & $-1.1^\circ$ & 73.5 \\
\bottomrule
\end{tabular}
\end{table}

It should be noted that a perfect volumetric correspondence between the reconstructed point clouds and the GCS model is not expected. The PRT provides the average location of the scattering plasma along the LOS and therefore cannot recover the depth extent of the entire CME \citep{Mierla:2009}. Consequently, the reconstructed point clouds should be regarded as sampling the 3D plasma distribution rather than reproducing the complete volume described by the GCS flux-rope model.
\section{Discussion and conclusions} \label{sec:conclusions}

In this work, we presented PRICM3, a new framework that transforms PRT topographical maps generated from multiple viewpoints into 3D point-cloud representations of CMEs within a common heliocentric reference frame. Rather than introducing a new reconstruction technique, PRICM3 provides a means of exploiting the 3D information encoded in PRT topographical maps. By converting these maps into filtered 3D point clouds, the framework enables independent reconstructions obtained from different coronagraphs and viewing geometries to be directly compared and quantitatively assessed using dedicated consistency estimators. It also facilitates comparisons with complementary geometrical models, such as the GCS model. The resulting point-cloud representation makes it easier to identify regions where independent reconstructions converge or diverge, providing a basis for evaluating their mutual consistency and for developing future quantitative metrics to assess the consistency and reliability of PRT reconstructions across different coronagraphs and viewing geometries.

The proof-of-concept application to the slow limb CME of 28 October 2021 demonstrates that independent PRT reconstructions obtained from Metis, LASCO C2, and STEREO-A/COR2 yield mutually consistent 3D CME localizations. The quantitative estimators introduced in this work show that all reconstructed cloud centroids remain within the uncertainty of the independently inferred GCS propagation direction, while the overlap analysis confirms substantial agreement with the GCS volume, reaching more than 80\% when considering the adopted reconstruction uncertainties. Furthermore, we would like to highlight that this filtering procedure is especially valuable because the discarded pixels, i.e., those found deviating from the main CME body, are not necessarily spurious. Instead, they may contain valuable information regarding the 3D location of other ambient coronal structures, thereby allowing for their independent physical interpretation. In this respect, PRICM3 enables advanced coronal diagnostics focusing on 3D GCS volume comparisons with shock-sheath or draping patterns and streamer compression.

The present implementation has been designed as a modular framework that can be readily extended to additional coronagraphs and future developments. In particular, the rapidly growing availability of polarimetric observations from instruments such as Metis, PROBA-3/ASPIICS, and PUNCH makes tools for the visualization, comparison, and quantitative assessment of independent 3D PRT reconstructions increasingly valuable.

In particular, during Solar Orbiter perihelion passages, Metis and ASPIICS will sample a similar heliocentric distance range in the inner corona from different viewing geometries. Combined with the high spatial resolution of both coronagraphs over this region, these observations will provide unique opportunities to investigate the 3D distribution of fine CME substructures with unprecedented detail. Such observations may prove particularly valuable for investigating localized polarization signatures associated with specific CME features, such as the leading-edge draping and sheath region \citep{McComas:1988}, potentially providing new insight into the relationship between the reconstructed plasma distribution and the underlying magnetic-field topology.

At the opposite end of the observational spectrum, PUNCH will represent an equally compelling application of PRICM3. Whereas Metis and ASPIICS will enable high-resolution studies of the inner corona over similar heliocentric distance ranges in the aforementioned observation conditions, the Wide Field Imager (WFI) onboard PUNCH will provide an unprecedented FOV extending to elongations of approximately $45^\circ$. 
Such observations will allow PRICM3 to be applied to
the 3D reconstruction of CMEs over much larger spatial
scales than those currently accessible by classical
coronagraphs, following an appropriate adaptation
of the geometrical framework to the large-elongation
observations.
%

As coordinated polarimetric observations continue to expand both in spatial coverage and observational capability, PRICM3 offers a versatile framework for exploiting their 3D diagnostic potential, paving the way toward more systematic analyses of CME evolution across multiple observing geometries.
\begin{acknowledgments}
Y.D.L. would like to thank Dr. M. Mierla for insightful discussions on the Polarization Ratio Technique and Dr. B. Hofer for valuable discussions on the statistical interpretation of the results.\\ 
Y.D.L. was supported by an INAF scholarship funded under the ASI-INAF agreement ``Supporto per la realizzazione degli strumenti Metis, SWA DPU e STIX'' (CUP F86C18000570005).\\
We acknowledge the data from SOHO, STEREO, and Solar Orbiter missions.
Solar Orbiter is a space mission of international collaboration between ESA and NASA, operated by ESA.  Metis was built and operated with funding from the Italian Space Agency (ASI), under contracts to the National Institute of Astrophysics (INAF) and industrial partners. Metis was built with hardware contributions from Germany (Bundesministerium für Wirtschaft und Energie through DLR), from the Czech Republic (PRODEX) and from ESA.  
Metis team thanks the former PI, Ester Antonucci, for leading the development of Metis until the final delivery to ESA.\\
This research was funded in whole or in part by the Austrian Science Fund (FWF) 10.55776/ESP2660425. For the purpose of open access, the author has applied a CC BY public copyright licence to any Author Accepted Manuscript version arising from this submission.\\
The authors acknowledge the financial support provided by the University of Graz.\\
This research is co-sponsored by the DynaSun project and has thus received funding under the Horizon Europe programme of the European Union under grant agreement (no. 101131534).\\
S.E.G. acknowledges the support of NASA Explorers program (Contract 80GSFC14C0014) and the National Center for Atmospheric Research, a major facility sponsored by the U.S. National Science Foundation under Cooperative Agreement No. 1852977.
\end{acknowledgments}
\begin{contribution}
Y.D.L. contributed to the conception of the work, development and management of the PRICM3 tool, data analysis, interpretation, and the drafting and revision of the manuscript. M.T. contributed to the conception of the work, data analysis, interpretation of results, and the drafting and revision of the manuscript. F.M.L. and H.C. contributed to the data analysis, interpretation of results, and the drafting and revision of the manuscript. S.E.G. and R.S. contributed to the interpretation of results, the drafting and revision of the manuscript. L.D.L. and M.R. contributed to the interpretation of results, and the revision of the manuscript. L.A., A.B., C.C., M.F., F.F., G.J., F.L., M.P., G.R., and C.S. contributed to the planning of the Metis observations acquired during the CME event under investigation and the revision of the manuscript.
\end{contribution}
%
\facilities{Solar Orbiter (Metis), SOHO (LASCO), STEREO-A (SECCHI/COR2)}

\software{Sunpy \citep{sunpy_community2020}, GCS Python implementation by C. Kay (private communication), IDL SolarSoftWare package (SSW; \citealt{Freeland:1998})}         


\appendix
\section{Derivation of PRT Topographical Maps}\label{appendix:a}
The PRT topographical maps used as input to PRICM3 were generated independently of the framework, following the methodology described in \citet{MoranDavila:2004,Moran:2010,Dere:2005,Zuccarello:2013}. The PRT exploits the polarization properties of Thomson-scattered WL emission \citep{Billings:1966} to infer the location of scattering plasma relative to the observer's POS.

\begin{figure}
\centering
\includegraphics[width=0.4\columnwidth]{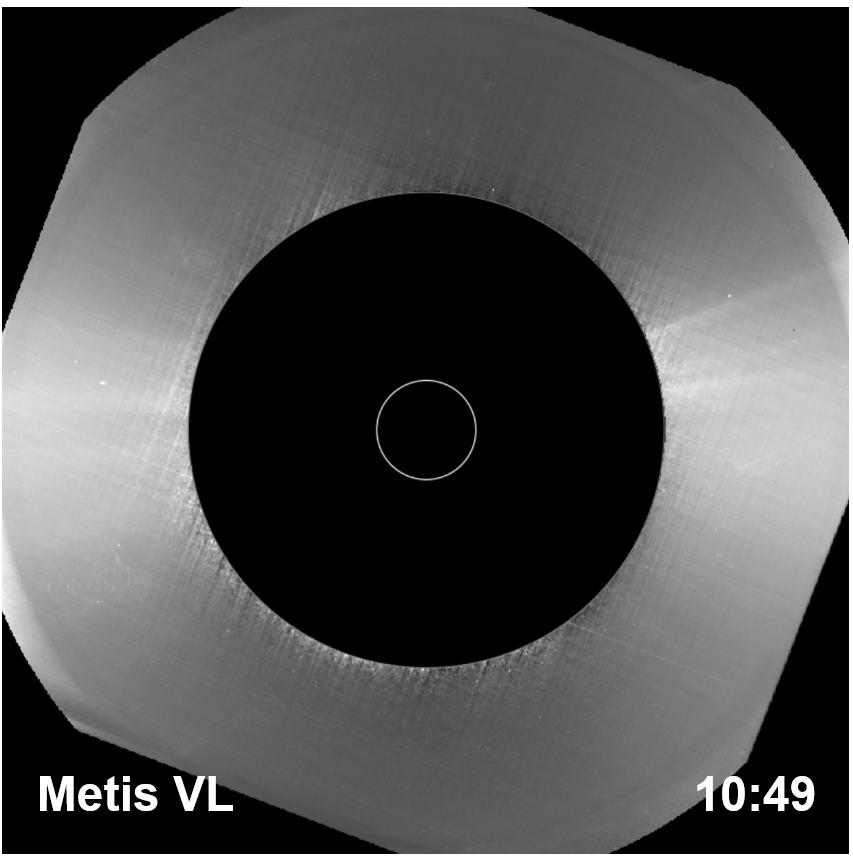}
\caption{Pre-event Metis tB image acquired on 28 October 2021 at 10:49 UT, used as a reference for the background subtraction. The image was enhanced using the Normalizing Radial Graded Filter \citep[NRGF;][]{Morgan:2006} technique to better highlight the coronal structures.}
\label{fig:pre-event}
\end{figure}

The method makes use of $pB$ and unpolarized-brightness ($uB=tB-pB$) images to construct the measured ratio $R_{\mathrm{m}}=pB/uB$, whose value depends on the position of the scattering electrons along the observer's LOS. To isolate the CME signal, the input data were prepared by subtracting the coronal background, with the aim of removing the F-corona contribution and coronal structures not associated with the CME. For the COR2-A observations, we first subtracted the standard monthly-minimum coronal background and subsequently applied a pre-event subtraction to isolate the CME signal. For the LASCO C2 and Metis observations, satisfactory results were obtained by subtracting only the pre-event images (see Fig. \ref{fig:pre-event}). In the case of Metis, no standard background image is officially distributed because of the significant variability of the observing conditions, including acquisition parameters and the portion of the corona sampled at different spacecraft heliocentric distances. Isolated CME signals are shown in the $tB$ panels from Fig.~\ref{fig:top_maps}.

Following background subtraction, an initial brightness-based filtering was applied to the polarized ($pB$) and unpolarized ($uB$) brightness images to exclude pixels with low signal levels (potentially affected by background noise) before computing the measured polarization ratio, $R_m$\footnote{Alternative filtering criteria have also been proposed for polarization-ratio analyses. For example, \citet{Moran:2010} discarded pixels whose polarization angle differed by more than $10^\circ$ from the expected tangential direction, thereby removing measurements with low signal-to-noise ratio. This criterion was not adopted in the present work.}. Instrument-dependent brightness thresholds were adopted, and $R_m$ was calculated only for pixels satisfying both the $pB$ and $uB$ selection criteria. The thresholds were deliberately chosen to be conservative to preserve faint CME signals and avoid excessively restricting the reconstructed structures. Consequently, residual contributions from the surrounding corona may still be present in the resulting topographical maps and are addressed through additional filtering during the subsequent analysis performed with PRICM3.

By comparing the measured ratio $R_{\mathrm{m}}$ with the theoretical polarization ratio $R_{\mathrm{t}}$ expected from Thomson-scattering geometry, a proxy for the distance of the scattering plasma from the POS can be determined. The theoretical values of $R_{\mathrm{t}}$ were computed using the SolarSoft routine \texttt{eltheory.pro}, following the approach described by \citet{Dere:2005}. The resulting quantity, here denoted as $|z|$, represents the average LOS-weighted distance of the emitting plasma from the POS, where $z$ is the coordinate along the observer's LOS. Because the reconstruction is based on integrated WL emission, the inferred distance corresponds to the average location of all scattering elements contributing to a given pixel (note, this assumes a sufficiently compact source; for a discussion of how LOS extent relative to POS projected radius introduces departures from such a center-of-mass average, see Sect. 2.5 of \citet{Gibson:2026}). Typical uncertainties in the reconstructed $|z|$ values have been estimated to range between approximately 1.5\% and 3.0\% under favorable observing conditions \citep{Lopez:2012,Zuccarello:2013}. Additional uncertainties may arise from LOS integration effects, departures from the assumption of a single dominant scattering structure, and contamination from unrelated coronal features. 
The intrinsic front/back ambiguity of the reconstructed
LOS coordinate can be constrained using complementary
EUV observations, which provide information on the CME
source-region location and propagation direction.
When the source region cannot be unambiguously
identified or is not visible, the front/back ambiguity
can also be constrained through independent GCS fitting
or by combining observations from multiple viewpoints.
The inferred distance from the POS, expressed in solar radii, is encoded in the color scale shown in the $|z|$-map panels of Fig.~\ref{fig:top_maps}. Such derived topographical maps constitute the input data products for PRICM3.

The application of the PRT to large-elongation
observations, such as those provided by PUNCH/WFI,
marks a transition to a different Thomson-scattering
regime.
Unlike conventional coronagraphs, PUNCH samples Thomson-scattered light over much larger heliocentric distances, where two assumptions commonly adopted for coronagraph observations require revision.
First, the finite angular extent of the solar disk becomes negligible, allowing the Sun to be treated as a point-like illumination source \citep[see][]{Hundhausen:1993}.
Under this approximation, the Van de Hulst coefficients disappear from the Thomson-scattering formalism \citep[see][]{DeForest:2013}, considerably simplifying the polarization-ratio calculations and eliminating the need to compare the measured ratio $R_{\mathrm{m}}$ with theoretical values, as described above. Conversely, at large elongations the assumption of parallel lines of sight ceases to be valid. The increasing separation between the observer's POS and the corresponding Thomson sphere must be explicitly taken into account when constructing the topographical maps, requiring an appropriate reformulation of the geometrical framework adopted by PRICM3. The modifications of the Thomson-scattering formalism required in this regime are discussed by \citet{Gibson:2026}.

\section{Additional comparison between the Metis PRT Reconstruction and the GCS model}\label{appendix:b}
\begin{figure}
\centering
\includegraphics[width=0.6\columnwidth]{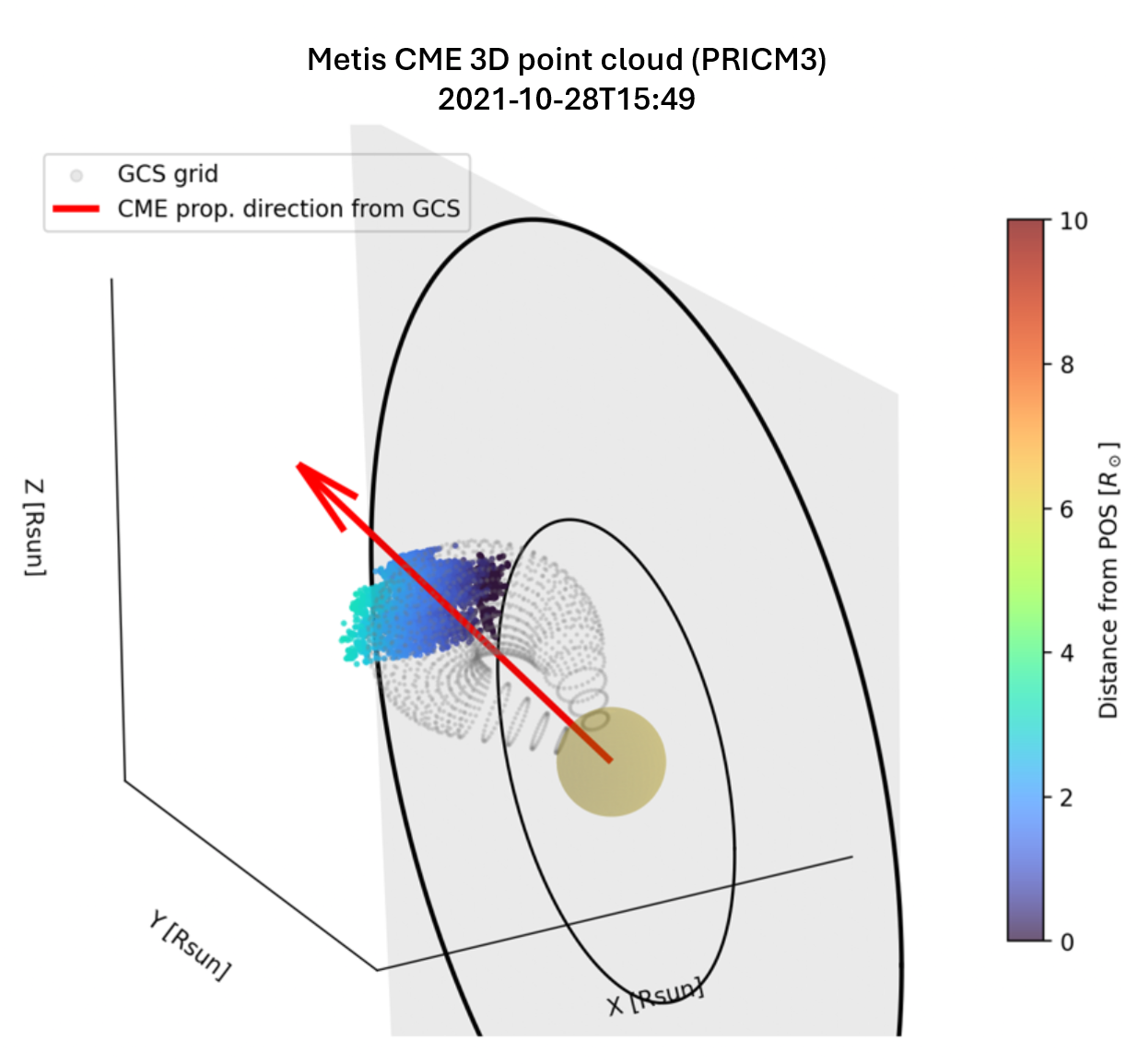}
\caption{Three-dimensional comparison between the PRT point cloud reconstructed with PRICM3 from the Metis observation at 15:49~UT on 28 October 2021 and the corresponding GCS model. The GCS apex height was interpolated to the Metis observation time, consistently with the procedure adopted for the comparisons presented in Fig.~\ref{fig:results}. Point colors indicate the distance from the POS.}
\label{fig:Metis_1549}
\end{figure}

Figure~\ref{fig:Metis_1549} presents an additional three-dimensional comparison between the PRT point cloud reconstructed from the Metis observation at 15:49~UT on 28 October 2021 and the corresponding GCS model. As in the comparisons presented in Fig.~\ref{fig:results}, the GCS geometry was reconstructed using the model parameters derived independently from forward modeling, with the apex height interpolated to the Metis observation time. The figure provides a complementary view of the reconstructed CME structure, allowing a direct visual assessment of the spatial correspondence between the Metis point cloud and the GCS envelope.
\bibliography{PRICM3_biblio}{}
\bibliographystyle{aasjournalv7.1}
%
%


\end{document}